\documentclass[twocolumn,showpacs,preprintnumbers,amsmath,amssymb]{revtex4}

\usepackage{graphicx}
\usepackage{dcolumn}
\usepackage{amsmath, bm}
\usepackage{amssymb}

\newcommand{ \bra }[1]{ \langle #1 | }
\newcommand{ \ket }[1]{ | #1 \rangle }

\newcommand{ \sH }{ \mathcal{H} }

\newcommand{ \sD }{ \mathcal{D} }
\newcommand{ \sL }{ \mathcal{L} }

\newcommand{ \sM }{ \mathcal{M} }

\newcommand{ \sE }{ \mathbb{E} } 

\newcommand{ \half }{ \frac{1}{2} }
\newcommand{ \eo }{ \varepsilon_0 }

\newcommand{ \Tr }{ \mbox{Tr} }

\begin{document}

\preprint{APS/123-QED}

\title{Continuous measurement feedback control of a Bose-Einstein condensate using phase contrast imaging}

\author{S. S. Szigeti}
\author{M. R. Hush}
\author{A. R. R. Carvalho}
\affiliation{Department of Quantum Science, Research School of Physics and Engineering, The Australian National University, ACT 0200, Australia}
\author{J. J. Hope}
\affiliation{Australian Centre for Quantum-Atom Optics, Department of Quantum Science, Research School of Physics and Engineering, The Australian National University, ACT 0200, Australia}

\date{\today}

\begin{abstract}
We consider the theory of feedback control of a Bose-Einstein condensate (BEC) confined in a harmonic trap under a continuous measurement constructed via non-destructive imaging. A filtering theory approach is used to derive a stochastic master equation (SME) for the system from a general Hamiltonian based upon system-bath coupling. Numerical solutions for this SME in the limit of a single atom show that the final steady state energy is dependent upon the measurement strength, the ratio of photon kinetic energy to atomic kinetic energy, and the feedback strength. Simulations indicate that for a weak measurement strength, feedback can be used to overcome heating introduced by the scattering of light, thereby allowing the atom to be driven towards the ground state. \end{abstract}

\pacs{03.75.Gg, 03.75.Pp, 05.40.Ca, 37.10.De}
\maketitle

\section{Introduction}
The atom laser is the most coherent source for atom optical experiments \cite{Ketterle_1997,Bouyer_1997}. However, the applicability of the atom laser as a tool for fundamental research is limited by noise that broadens the linewidth. This noise is due, in part, to instability in the spatial mode of the Bose-Einstein condensate (BEC) from which the atom laser beam is outcoupled. Excitations of the BEC spatial mode exist under general preparation conditions \cite{Drummond_Corney_1999}, and are generated when the atom laser is continuously pumped \cite{Haine_2004, Johnsson_Haine_Hope_2005}. One promising solution to this difficulty is to drive the BEC towards a stable spatial mode via the use of measurement feedback control. A control scheme based on a measurement of position has been shown to successfully cool a single atom in a harmonic trap close to the ground state from any initial state \cite{Wilson_2007}. However, although this control scheme can be engineered by placing the atom in a cavity \cite{Doherty_Jacobs_1999}, it is unclear whether it could be generalised to a many-atom BEC.  In this paper we consider a feedback control scheme based upon dispersive imaging, a technique that has already been implemented in multiple BEC laboratories.

Feedback on an atom laser was first applied in a single mode model to reduce phase noise caused by the interactions \cite{Wiseman_Thomsen_2001,Thomsen_Wiseman_2002}.  Improving the modal stability of a BEC using feedback control was first examined using a semiclassical model \cite{Haine_2004}, where it was shown that the system could be stabilised with feedback by modification of the trapping potential and the condensate's nonlinearity. This feedback scheme was then applied to a more realistic model of an atom laser that included pumping, damping and outcoupling \cite{Johnsson_Haine_Hope_2005}. While the semiclassical approximation allowed for an examination of the multimode behaviour of the atom laser, the effect of coupling the system to a measuring device was ignored. The measurement backaction was included in a model of a trapped single particle considered by Doherty and Jacobs 
\cite{Doherty_Jacobs_1999}, who showed how a position measurement arises from placing an atom in a cavity, and solved the optimal control problem for an initial Gaussian state. It was later shown that the filter equation could be solved, and the atom could be cooled from any arbitrary state \cite{Wilson_2007}. The position measurement relied upon the assumption that the atom is trapped in a region small compared to the wavelength of light within the cavity. However this assumption is false for a modestly sized BEC in an optical cavity. Moreover, even if this type of position measurement could be engineered for a BEC in theory, either by situating the BEC in a cavity or otherwise, this does not imply that it would be easy to implement in practice. Indeed, a weak position measurement of a condensate has not been experimentally realised. There is thus a clear preference towards developing a feedback scheme that uses a well-established technique of measurement.

There are two commonly used techniques for measuring a BEC of alkali atoms. The first method, termed \emph{absorption imaging} \cite{Hau_1998}, shines near-resonant laser light on the condensate. Those photons which interact with atoms from the condensate are absorbed, leaving a `shadow' which can be detected using an array of CCD cameras. Thus absorption imaging gives a measurement of the column density (the number density integrated along the line-of-sight of the laser) of the BEC, which gives information about the spatial distribution of atoms in the condensate. The key advantage of absorption imaging is that any data obtained is independent of the intensity of the light, the time of exposure and many properties of the CCD array. A downside, however, is that the absorption of photons heats the atoms sufficiently to destroy the BEC. 

\emph{Phase-contrast imaging} \cite{Ketterle_1996, Bradley_Sackett_Hulet_1997} is an alternative method of imaging that uses light highly detuned from resonance. The interaction of the BEC with the light gives a phase profile for the condensate, which can be used to reconstruct the density profile. In principle, the detuning can be sufficiently large such that a measurement of the density is minimally destructive.  The theoretical limits for the signal generated from this dispersive measurement scheme are not in fact different from absorption imaging in the case of low optical depth \cite{Lye_2003,Hope_2004,Hope_2005}. However, for a BEC, it is far easier to achieve near optimal sensitivity for a given level of spontaneous emission in a dispersive measurement. Experiments showing successive repeated measurements with phase-contrast imaging \cite{Ketterle_1996} suggest that it may be possible to use this measurement technique non-destructively on the timescale needed to perform the feedback control.  In this paper we show that measurements of a BEC via the technique of phase-contrast imaging can be used to construct a feedback control scheme.  

In Sec.~\ref{model} we present our model of the system and the associated stochastic master equation (SME) for the quantum filter. After adiabatically eliminating the excited state we recapture the SME considered by Dalvit \emph{et al.} in the limit where the size of the condensate is much larger than the wavelength of light \cite{Dalvit_2002}. The form of feedback, which is modelled by the inclusion of a control Hamiltonian, is also outlined. In Sec.~\ref{simulation} the quantum filter is numerically solved in the limit where the atomic sample contains only a single atom. The dependence of the atom's final steady state energy on the measurement strength, ratio of photon recoil energy to typical atomic kinetic energy and form of feedback is discussed. Attempts at a numerical solution for the quantum filter under the semiclassical limit are also elucidated.

\section{Model and Filter Derivation \label{model}}
\begin{figure}[ht!]
\centering
\includegraphics[scale=0.6]{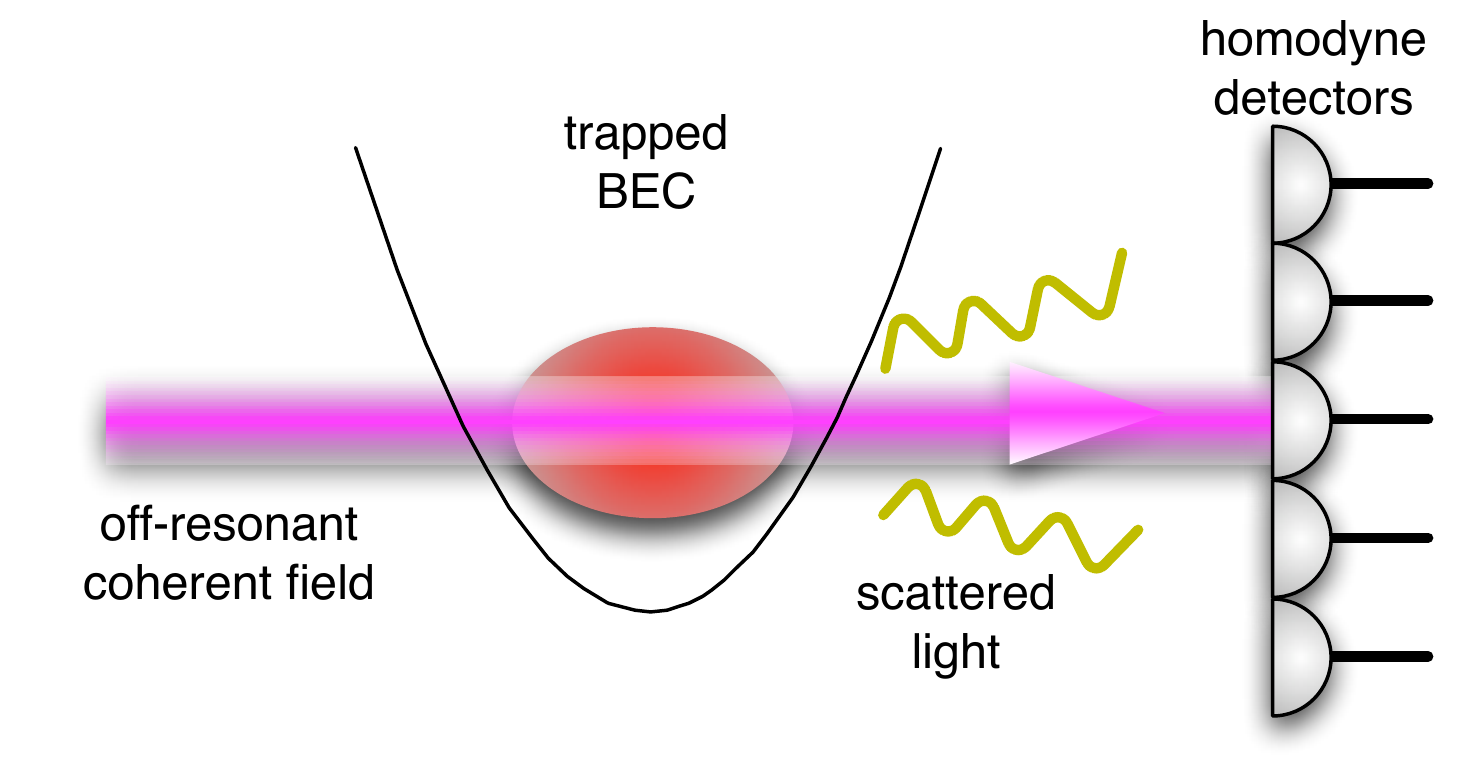}
\caption{\label{BEC_diagram} (Colour online) Diagram illustrating how one could perform a non-destructive density measurement on a BEC. The BEC is illuminated with highly off-resonant laser light. The interaction between the field and the atoms is registered as a phase shift on any light scattered from the condensate. Such a phase shift is measurable by homodyne detection in the phase quadrature.}
\end{figure}
The system under analysis is a BEC magnetically confined in a harmonic trap and illuminated with an off-resonant coherent field directed along the z-direction (see Fig. \ref{BEC_diagram}). Light scattered from the condensate is detected by an array of homodyne detectors. The total Hamiltonian for the combined system is
\begin{equation}
	\hat{H}_{\text{tot}} = \hat{H}_\text{sys} + \hat{H}_{\text{B}} +  \hat{H}_{I}. \label{H_tot}
\end{equation}
For simplicity, it is assumed that the atoms have been configured such that a transition between only two states is possible. These levels will be labelled the ground ($g$) and excited ($e$) states. Under this assumption, the system Hamiltonian can be written as 
\begin{align}
	\hat{H}_\text{sys} &= \int d^3\bm{r} \,\hat{\psi}_g^\dag(\bm{r})H_a(\bm{r})\hat{\psi}_g(\bm{r}) \notag \\
					& \qquad + \int d^3\bm{r} \,\hat{\psi}_e^\dag(\bm{r})\left(H_a(\bm{r}) + \hbar \omega_0\right)\hat{\psi}_e(\bm{r}),
\end{align}
where $\omega_0$ is the resonant frequency of the atoms and $\hat{\psi}_i(\bm{r})$ is the field operator that annihilates a boson from the $i$th atomic level at position $\bm{r}$, obeying the commutation relation $[\hat{\psi}_i(\bm{r}), \hat{\psi}_j^\dag(\bm{r'})] = \delta_{ij}\delta^3(\bm{r} - \bm{r'})$.  $H_a(\bm{r})$ is the single particle Hamiltonian 
\begin{equation}
H_a(\bm{r}) = -\frac{\hbar^2}{2m}\nabla^2 + \half m \omega^2 \bm{r}^2 + H_\text{control}(\bm{r}).
\end{equation}
For simplicity, we have taken the interaction energy between atoms to be negligible.  $H_\text{control}(\bm{r})$ is the single particle control Hamiltonian, which we will specify shortly. The Hamiltonian for the electromagnetic field is 
\begin{equation}
H_B = \int d^3 \bm{p}\, \hbar c |\bm{p}| \left(\hat{a}_{\bm{\varepsilon}}^\dag(\bm{p})\hat{a}_{\bm{\varepsilon}}(\bm{p}) + \hat{a}_{\bm{\varepsilon}'}^\dag(\bm{p})\hat{a}_{\bm{\varepsilon}'}(\bm{p})\right),
\end{equation}
where $\bm{\varepsilon}(\bm{p})$ and $\bm{\varepsilon}'(\bm{p})$ are the two unit vectors required to describe the polarisation of the field, $\bm{p} = (k_x,k_y,k_z)$, and $[\hat{a}_{i}(\bm{p}),\hat{a}_{j}^\dag(\bm{p}')] = \delta_{ij}\delta^3(\bm{p} - \bm{p'})$. It can be assumed that the wavelength of the driving laser is much greater than the Bohr radius of an atom. Thus the interaction between the electric field and an atom can be adequately modelled by approximating the atom as a dipole. In this case the most general interaction Hamiltonian $\hat{H}_I$ for a field of two-level bosonic atoms interacting with the electromagnetic field is \cite{Meystre_2001}
\begin{align}
	\hat{H}_I &= - \int d^3 \bm{r}  \, \hat{\psi}^\dag_g(\bm{r}) \left[ \bm{d}_{eg} \cdot \hat{\bm{E}}(\bm{r},t) \right] \hat{\psi}_e(\bm{r}) \notag \\ 
			&	\qquad \quad - \int d^3 \bm{r}  \, \hat{\psi}^\dag_e(\bm{r}) \left[ \bm{d}_{ge} \cdot \hat{\bm{E}}(\bm{r},t) \right] \hat{\psi}_g(\bm{r}). \label{general_int}
\end{align}
where $\bm{d}_{ij}$ is the transition electric dipole vector between the $i$th and $j$th level of the atom, defined as
\begin{equation}
 	\bm{d}_{ij} = \bra{i} \bm{r}\ket{j},
\end{equation} 
where $\ket{i}$ denotes the state vector for the $i$th level. We choose the phase of the internal states such that $\bm{d}_{eg}$ is real, which implies that $\bm{d}_{eg} = \bm{d}_{ge}$. $\hat{\bm{E}}$ is the quantised electric field operator, which can be expressed in terms of a basis of creation and annihilation operators as \cite{Gardiner_Zoller_2004}
\begin{equation}
	\hat{\bm{E}}(\bm{r}) = i\int d^3 \bm{p} \, \sum_{\bm{\epsilon}} \sqrt{\frac{\hbar \omega(\bm{p}) }{2(2\pi)^3 \eo}}\left\{ \hat{a}_{\bm{\epsilon}}(\bm{p})e^{i\bm{p} \cdot \bm{r}} - h.c. \right\}\bm{\epsilon}(\bm{p}).
\end{equation}

Importantly, the coupling of the system to the bath allows us to make a number of standard quantum optical and reservoir approximations, thereby leading to tractable equations. We demonstrate in appendix~\ref{appendix_derivation} that making such approximations yields the following Ito quantum stochastic differential equation (QSDE) for the unitary of the system and bath:
\begin{align}
	d\hat{U}(t) 	&= \left\{\left(-\frac{i}{\hbar}\hat{H} - \half\int dx \, \hat{M}^\dag(x)\hat{M}(x) \right) dt  \right.  \notag \\
			& \quad \left.+ \int dx \, \left(\hat{M}(x) d\hat{A}^\dag(x,t) - \hat{M}^\dag(x) d\hat{A}(x,t)\right)\right\}\hat{U}(t), \label{dU}
\end{align}
where
\begin{align}
\hat{H} 	&= \int dx \left\{ \hat{\psi}_g^\dag(x)H_a(x)\hat{\psi}_g(x) + \hat{\psi}_e^\dag(x)\left(H_a(x) + \hbar \Delta \right)\hat{\psi}_e(x)\right\} \notag\\
			& \, + i \sqrt{\frac{F_0 k_0^2 \hbar}{2(2\pi)^3 \eo}}  [ \bm{d}_{ge} \cdot \bm{\epsilon}(\bm{0})]\int dx \, \left( \hat{\psi}^\dag_g(x)\hat{\psi}_e(x) - h.c. \right), \label{H_f}
\end{align}
and
\begin{align*}
	\hat{M}(x)		&= \int dx' \hat{\psi}_g^\dag(x') \hat{\psi}_e(x')\Gamma(x - x'), \\
	\Gamma(x)	&= \int_{-k_0}^{k_0} dk_x \sqrt{\frac{\gamma(k_x)}{2\pi}}e^{ik_x x}, \\
	\gamma(k_x)	&= \frac{k_0^2}{8\hbar \eo} \int_{-\sqrt{k_0^2 - k_x^2}}^{\sqrt{k_0^2 - k_x^2}} dk_y [ \bm{d}_{ge} \cdot \bm{\epsilon}]^2 \frac{e^{ -z_0^2 \left[ (k_0 - k_z)^2 + k_y^2\right]/2}}{\sqrt{k_0 ^2 - k_x^2 - k_y^2}},
\end{align*}
for detuning $\Delta = \omega_0 - \omega_L$ and photon flux of the laser $F_0$. $z_0$ is the characteristic length scale of the condensate in the tightly trapped $z$ and $y$ directions (i.e. $z_0 = \sqrt{\hbar/m\omega_z}$). The quantum Wiener increment $d\hat{A}(x,t)$ obeys the Ito differential rule $d\hat{A}(x,t)d\hat{A}^\dag(x',t) = \delta^2(x - x')dt$ \cite{Gardiner_Zoller_2004}. This term models the noise introduced into the system due to vacuum fluctuations in the bath. The inner product between the dipole and polarization vectors can be shown to equal \cite{Mandel_Wolf_1995}
\begin{equation}
	[ \bm{d}_{ge} \cdot \bm{\epsilon}(\bm{p})]^2 = |\bm{d}_{ge}|^2 - [\bm{p} \cdot \bm{d}_{ge}]^2/|\bm{p}|^2.
\end{equation}
The above QSDE corresponds to a `cigar'-shaped condensate where the BEC is tightly confined in the $z$ and $y$ dimensions. We present the one-dimensional version of the unitary evolution simply because one dimensional simulations require less computuational power. In appendix~\ref{appendix_derivation} a more general two-dimensional derivation for a `pancake' shaped condensate is given. 

If we perform a homodyne measurement of the phase quadrature on the laser light after it has interacted with the atoms, then it can be shown that the best-estimate (in the least-squares sense) $\pi_t$ of any system observable $\hat{X}$ is given by the equation \cite{van_Handel_2005, Bouten_van_Handel_James_2007}
\begin{align}
	d\pi_t(\hat{X}) 	&= \pi_t(\sL[\hat{X}]) dt -i\int dx \, \left\{ \pi_t\left(\hat{X}\hat{M}(x) - \hat{M}^\dag(x) \hat{X}\right) \right.\notag \\
				& \quad \left. - \pi_t(\hat{X}) \pi_t\left(\hat{M}(x) - \hat{M}^\dag(x)\right) \right\}dW(x,t), \label{filter_eqn}
\end{align}
where we have defined the Lindblad generator
\begin{align}
	\sL[\hat{X}] &= \frac{i}{\hbar}[\hat{H}, \hat{X}] + \int dx \left(\hat{M}^\dag(x) \hat{X} \hat{M}(x) \right. \notag \\
				& \qquad \left. - \half \{ \hat{M}^\dag(x) \hat{M}(x), \hat{X}\}\right).
\end{align}
$dW(x,t)$ is the classical Wiener increment (i.e. Gaussian white noise). It satisfies $dW(x,t)dW(x',t) = \delta(x-x') dt$. This noise is the random wavefunction collapse that corrupts the homodyne measurement signal. By defining the conditional density operator $\hat{\rho}_c$ by $\pi_t(\hat{X}) = \Tr[\hat{X}\hat{\rho}_c]$ we can construct the following stochastic master equation (SME) for the quantum filter:
\begin{align}
	d\hat{\rho}_c 	&= -\frac{i}{\hbar}[\hat{H},\hat{\rho}_c] dt + \int dx \, \sD[-i\hat{M}(x)] \hat{\rho}_c dt \notag \\
				& \qquad	+ \int dx \sH[-i\hat{M}(x)] \hat{\rho}_c dW(x,t),\label{rho_c}	
\end{align}
where 
\begin{align*}
	\sD[\hat{c}]\hat{\rho}	&= \hat{c}\hat{\rho} \hat{c}^\dag -  \half\{ \hat{c}^\dag \hat{c}, \hat{\rho}\}, \\
	\sH[\hat{c}]\hat{\rho}	&= \hat{c}\hat{\rho} + \hat{\rho} \hat{c}^\dag - \Tr[(\hat{c}+\hat{c}^\dag)\hat{\rho}]\hat{\rho},
\end{align*}
for any arbitrary operator $\hat{c}$. 

\subsection{Adiabatic elimination}
For the trapped BEC under consideration the detuning of the laser is necessarily large. This allows us to adiabatically eliminate the excited state. We first transform to the following dimensionless units: 
\begin{equation}
	\xi = \frac{x}{x_0}, \qquad \kappa_i = \frac{k_i}{k_0}, \qquad \tau' = t \Delta ,
\end{equation}
where $x_0 = \sqrt{\hbar/m\omega_T}$ is the characteristic length scale of the trap in the $x$ direction, which has trapping frequency $\omega_T$. The field operators are made dimensionless by the transformation $\hat{\psi}_i(x) \to \hat{\psi}_i(\xi)/\sqrt{x_0}$. Define the following parameters:
$$ \eta = k_0 x_0, \quad \Gamma_\text{sp} = \frac{1}{4\pi \eo}\frac{4d_{ge}^2k_0^3}{3\hbar}, \quad \Omega = \frac{d_{ge}}{\hbar}\sqrt{\frac{F_0 k_0 \hbar}{2(2\pi)^3 \eo}},$$
where $d_{ge} = |\bm{d}_{ge}|$. Then Hamiltonian (\ref{H_f}) and conditional master equation (\ref{rho_c}) can be written as
\begin{align}
	&\hat{H} 	= \frac{\omega_T}{\Delta} \int d\xi \left( \hat{\psi}_g^\dag(\xi)H_a(\xi)\hat{\psi}_g(\xi) + \hat{\psi}_e^\dag(\xi)H_a(\xi)\hat{\psi}_e(\xi)\right) \notag \\
		& + \int d\xi \, \hat{\psi}_e^\dag(\xi)\hat{\psi}_e(\xi)+ i \frac{\Omega}{\Delta} \int d\xi \, \left( \hat{\psi}^\dag_g(\xi)\hat{\psi}_e(\xi) - \hat{\psi}^\dag_e(\xi)\hat{\psi}_g(\xi) \right) \label{H_dim} 
\end{align}
and
\begin{align}
d\hat{\rho}_c 	&= -i[\hat{H},\hat{\rho}_c] d\tau' + \frac{\Gamma_\text{sp}}{\Delta} \int d\xi \, \sD[\hat{\sM}(\xi)] \hat{\rho}_c d\tau' \notag \\
			& \qquad + \sqrt{\frac{\Gamma_\text{sp}}{\Delta}} \int d\xi \sH[\hat{\sM}(\xi)] \hat{\rho}_c dW(\xi,\tau'), \label{rho_c_dim}
\end{align}
respectively, where
\begin{align*}
	H_a(\xi) 	&= H_a(x)/\hbar\omega_T \\
	\hat{\sM}(\xi)	&= -i\int d\xi' \, \hat{\psi}_g^\dag(\xi') \hat{\psi}_e(\xi') \tilde{\Gamma}_\eta(\xi - \xi'), \\
	\tilde{\Gamma}_\eta(\xi)	&= \sqrt{\frac{3\eta}{8\pi}}\int_{-1}^1 d\kappa_x \, \sqrt{\frac{\tilde{\gamma}(\kappa_x)}{2\pi}} e^{i\eta \kappa_x \xi}, \\
	\tilde{\gamma}(\kappa_x) &= \int_{-\sqrt{1-\kappa_x^2}}^{\sqrt{1-\kappa_x^2}}d\kappa_y \,\frac{[ \bm{d}_{ge} \cdot \bm{\epsilon}]^2}{d_{ge}^2} \frac{e^{ -\frac{1}{4}w \left[ (1 - \kappa_z)^2 + \kappa_y^2\right]}}{\sqrt{1 - \kappa_x^2 - \kappa_y^2}},
\end{align*}
for $w = k_0^2 z_0^2/2$. Note that the Wiener increment has been rescaled by mapping $dW(x,t) \to dW(\xi,\tau')/\sqrt{x_0\Delta}$. This preserves the $\delta$-correlation $dW(\xi,\tau')dW(\xi',\tau') = \delta(\xi - \xi') d\tau'$.  We interpret $\Gamma_\text{sp}$ as the rate at which a single atom spontaneously emits into the bath, and $\Omega$ as the Rabi frequency for the atomic system. We assume that the detuning $\Delta$ of the laser is much larger than other characteristic frequencies in the system, i.e. $\Delta \gg \Omega, \Gamma_\text{sp}, \omega_T$. Furthermore, the intensity of the laser is sufficiently large such that $\Omega \gg \Gamma_\text{sp}, \omega_T$.

We begin the adiabatic elimination by calculating the evolution of $\hat{\psi}_e$, which can be found from eq.~(\ref{filter_eqn}) by using the Heisenberg equation:
\begin{align}
	\frac{d\hat{\psi}_e}{d\tau'}	&\approx -i\frac{\omega_T}{\Delta}[\hat{\psi}_e(\xi), \hat{H}].
\end{align}
Note that the terms proportional to $\Gamma_\text{sp}/\Delta$ and $\sqrt{\Gamma_\text{sp}/\Delta}$ are small and have thus been neglected. Furthermore, those terms in $\hat{H}$ proportional to $\omega_T/\Delta$ can be ignored, as they are small compared with the terms proportional to $\Omega/\Delta$ and unity. Thus
\begin{align}
	\frac{d\hat{\psi}_e(\xi)}{d\tilde{\tau}}	&\approx -i\frac{\omega_T}{\Delta}\left[\hat{\psi}_e(\xi),\int d\xi' \,\hat{\psi}_e^\dag(\xi')\frac{\Delta}{\omega_T}\hat{\psi}_e(\xi')\right] \notag \\ 
	& \qquad -i\frac{\omega_T}{\Delta}\left[\hat{\psi}_e(\xi),i \frac{\Omega}{\omega_T} \int d\xi' \, \left( \hat{\psi}^\dag_g(\xi')\hat{\psi}_e(\xi') - h.c. \right)\right]  \notag\\
	&= -i\hat{\psi}_e(\xi) - \frac{\Omega}{\Delta}\hat{\psi}_g(\xi). \label{dpsi_e_ad}
\end{align}
For large detuning, any atom excited by the laser spends a relatively short amount of time in the excited state before returning to the ground state. Moreover, there are very few atoms in the excited state in comparison to the ground state. Hence, on longer timescales it will appear that the population of excited atoms is tiny and changes very little. Thus, on this slower timescale we can approximate $d\hat{\psi}_e/d\tilde{\tau} \approx 0$. After making this approximation, eq.~(\ref{dpsi_e_ad}) gives
\begin{equation}
	\hat{\psi}_e(\xi) \approx i\frac{\Omega}{\Delta}\hat{\psi}_g(\xi).\label{psi_e_ad}
\end{equation}
As $\hat{\psi}^\dag_g$ and $\hat{\psi}_g$ do not commute, there is an ordering ambiguity upon substituting eq.~(\ref{psi_e_ad}) into $\hat{H}$. However, only one possible ordering yields a valid master equation:
\begin{align}
	\hat{H} 	&= \int d\xi \,\hat{\psi}_g^\dag(\xi){H}_0(\xi)\hat{\psi}_g(\xi) + \frac{\Omega^2}{\Delta^2}\int d\xi \,\hat{\psi}_g^\dag(\xi){H}_0(\xi)\hat{\psi}_g(\xi) \notag \\
		&\approx \int d\xi \,\hat{\psi}_g^\dag(\xi)H_a(\xi)\hat{\psi}_g(\xi), \label{H_ad}
\end{align}
where the term proportional to $1/\Delta^2$ is very small, and has hence been ignored. The conditional master equation simplifies to
\begin{align}
	d\hat{\rho}_c &= -i[\hat{H},\hat{\rho}_c] d\tau + \alpha \int d\xi \, \sD[\hat{\sM}_a(\xi)] \hat{\rho}_c d\tau \notag \\
				& \qquad + \sqrt{\alpha} \int d\xi \sH[\hat{\sM}_a(\xi)] \hat{\rho}_c dW(\xi,\tau), \label{drho_c_density}
\end{align}
where we have chosen the more convenient time scaling $\tau = \omega_T t$ and
\begin{align}
	\hat{\sM}_a(\xi) &= \int d\xi' \, \hat{\psi}_g^\dag(\xi') \hat{\psi}_g(\xi') \tilde{\Gamma}_\eta(\xi - \xi') \notag \\
	\alpha &= \frac{\Gamma_\text{sp}}{\omega_T}\frac{\Omega^2}{\Delta^2}.
\end{align}

It is now clear that there are two key dimensionless parameters upon which the system depends. The parameter $\alpha$ represents the strength of the measurement. For a larger $\alpha$, more information is obtained in a fixed time. However, there is also more measurement backaction due to the linear scaling with $\alpha$ of the decoherence. $\eta$ is the Lamb-Dicke parameter, which is clear when it is written as $ \eta = 2\pi x_0/\lambda = \hbar k_0/p_0$, where $\lambda$ is the wavelength of light emitted from the atoms and $p_0 = \sqrt{\hbar \omega m}$ is the characteristic momentum spread in the harmonic trap. Thus $\eta$ is proportional to the relative momentum `kick' an atom gets from scattering a photon. 

As a final note, typical experiments of the kind described above operate in the regime where $z_0 \gg \lambda$ (i.e. $w \gg 1$). Thus the exponential $\exp[-w(1-\kappa_z)^2]$, which appears in the integrand in the definition of $\tilde{\gamma}(\kappa_x)$, is tightly peaked about $\kappa_x = \kappa_y = 0$. Physically, this is indicative of the fact that photons are strongly scattered in the $z$ direction, and therefore have very little momentum in the $x$ and $y$ directions. Hence we are justified in expanding $\kappa_z = \sqrt{1 - (\kappa_x^2 + \kappa_y^2) }$ to second order within the exponent:
\begin{equation}
	\kappa_z \approx 1 - \frac{(\kappa_x^2 + \kappa_y^2)}{2k_0^2} - \frac{(\kappa_x^2 + \kappa_y^2)^2}{8k_0^4},
\end{equation}
in which case $(\kappa_z - k_0)^2 \approx (\kappa_x^2 + \kappa_y^2)^2/4k_0^2$. Furthermore, we can approximate $\kappa_z^{-1/2} \approx 1$ and $[\bm{d}_{ge} \cdot \varepsilon (\bm{p})] \approx d_{ge}$ with very little effect on the form of $\tilde{\gamma}(\kappa_x)$. Finally, we can extend the limits of integration over $\kappa_y$ (and, incidentally, those for integral over $\kappa_x$ which defines $\tilde{\Gamma}(\xi)$) to $\pm \infty$, to give a much simpler functional form for $\tilde{\gamma}$:
\begin{equation}
	\tilde{\gamma}(\kappa_x) = \int_{-\infty}^\infty d\kappa_y \, e^{-w \kappa_y^2}\exp \left[ -w(\kappa_x^2 + \kappa_y^2)^2/4\right].
\end{equation}
In this regime our result is in agreement with the master equation considered in \cite{Dalvit_2002}.

\subsection{Control}
Up to this point we have left the form of control completely arbitrary. Indeed this is the advantage of separating the control problem into estimation and control stages. Since the filter provides a best estimate of the system state, these control terms can be a function of any system observable. Thus one is free to concentrate on the design of an effective feedback scheme, secure in the knowledge that the filter for the system is independent of the choice of feedback. We consider a general control Hamiltonian of the form
\begin{equation}
	H_{\text{control}}(\xi) = \sum_{n=1}^\infty u_n(t) \xi^n. \label{control_H}
\end{equation}
Haine \emph{et al.}~\cite{Haine_2004} performed a semiclassical analysis that showed that the change in energy of the BEC was always non-negative for the choice
\begin{equation}
	u_n(t) = c_n\frac{d\left< \hat{x}^n\right>}{dt} = c_n\frac{n}{2}\left< \hat{p}\hat{x}^{n-1} + \hat{x}^{n-1}\hat{p}\right>, 
\end{equation}
for positive constants $c_n$. In this paper we will be primarily concerned with a feedback control consisting of only the first term in eq.~(\ref{control_H}). More precisely
\begin{equation}
	H_{\text{control}}(\xi) = c_1 \left< \hat{p} \right>\xi.
\end{equation}
A feedback control of this form was used in \cite{Wilson_2007} and \cite{Doherty_Jacobs_1999}. Physically, this control represents an adjustment of the trap minimum such that the motion of the atom is dampened. Such a control could be implemented experimentally via the use of changing magnetic fields.

\section{Simulation \label{simulation}}
\subsection{Single atom limit}
Important insight into the behaviour of a BEC under the above mentioned control scheme can be gathered by considering the physical limit of a single atom. In this case the density operator can be written as
\begin{equation*}
	\hat{\varrho} = \int d\xi \int d\xi' \, \varrho(\xi,\xi')\hat{\psi}^\dag(\xi)\ket{0}\bra{0}\hat{\psi}(\xi'),
\end{equation*}
where the coefficients $\varrho(\xi,\xi')$ are given by
\begin{equation*}
	\varrho(\xi,\xi') = \bra{0}\hat{\psi}(\xi) \hat{\rho} \hat{\psi}^\dag(\xi)\ket{0}.
\end{equation*}
The evolution of these coefficients is given by
\begin{equation}
	d\varrho(\xi,\xi') = \bra{0}\hat{\psi}(\xi)d \hat{\rho} \hat{\psi}^\dag(\xi)\ket{0}. \label{drho_single_coeff}
\end{equation}
Substituting eq.~(\ref{rho_c}) into eq.~(\ref{drho_single_coeff}) yields the following SME:
\begin{align}
	d\hat{\varrho}_c &= -i\left[ \hat{H}_a, \hat{\varrho}_c \right] d\tau + \tilde\alpha \int d\kappa_x \, \sD\left[\sqrt{\tilde\gamma(\kappa_x)} e^{-ik_0 \kappa_x\hat{x}} \right]\hat{\varrho}_c d\tau \notag \\ 
				&\qquad + \sqrt{\tilde\alpha}\int d\kappa_x \, \sH\left[\sqrt{\tilde\gamma(\kappa_x)}e^{-ik_0 \kappa_x \hat{x}}d\overline{W}^*(\kappa_x,\tau) \right]\hat{\varrho}_c, \label{rho_c_single}
\end{align}
where 
\begin{align*}
	\hat{H}_a	&= \half\left( \hat{p}^2 + \hat{x}^2\right) + c_1 \left< \hat{p} \right> \hat{x} \\
	\tilde{\alpha} &= \frac{3\alpha}{2\pi^2} \\
	d\overline{W}(\kappa_x,\tau)	&= \frac{1}{\sqrt{2\pi}} \int d\xi \, e^{-i\kappa_x \xi} dW(\xi,\tau).
\end{align*}
$d\overline{W}(\kappa_x,\tau)$ is the Fourier transform of the Wiener increment. It behaves somewhat differently to the traditional Wiener process: 
\begin{align*}
	d\overline{W}^{*}(\kappa_x,\tau) d\overline{W}(\kappa_x',\tau) &= \delta(\kappa_x - \kappa_x') d\tau \\
	d\overline{W}(\kappa_x,\tau) d\overline{W}(\kappa_x',\tau) &= \delta(\kappa_x + \kappa_x') d\tau.
\end{align*}
It is possible to express $d\overline{W}(\kappa_x,\tau)$ in terms of the Wiener increment $dW(\kappa_x,\tau)$. In particular,
\begin{equation}
	d\overline{W}(\kappa_x,\tau) = \half(i - 1)\left(dW(\kappa_x,\tau) + idW(-\kappa_x,\tau)\right). \label{lin_com_noise}
\end{equation}

\begin{figure}[t!]
\centering
\includegraphics[scale=0.55]{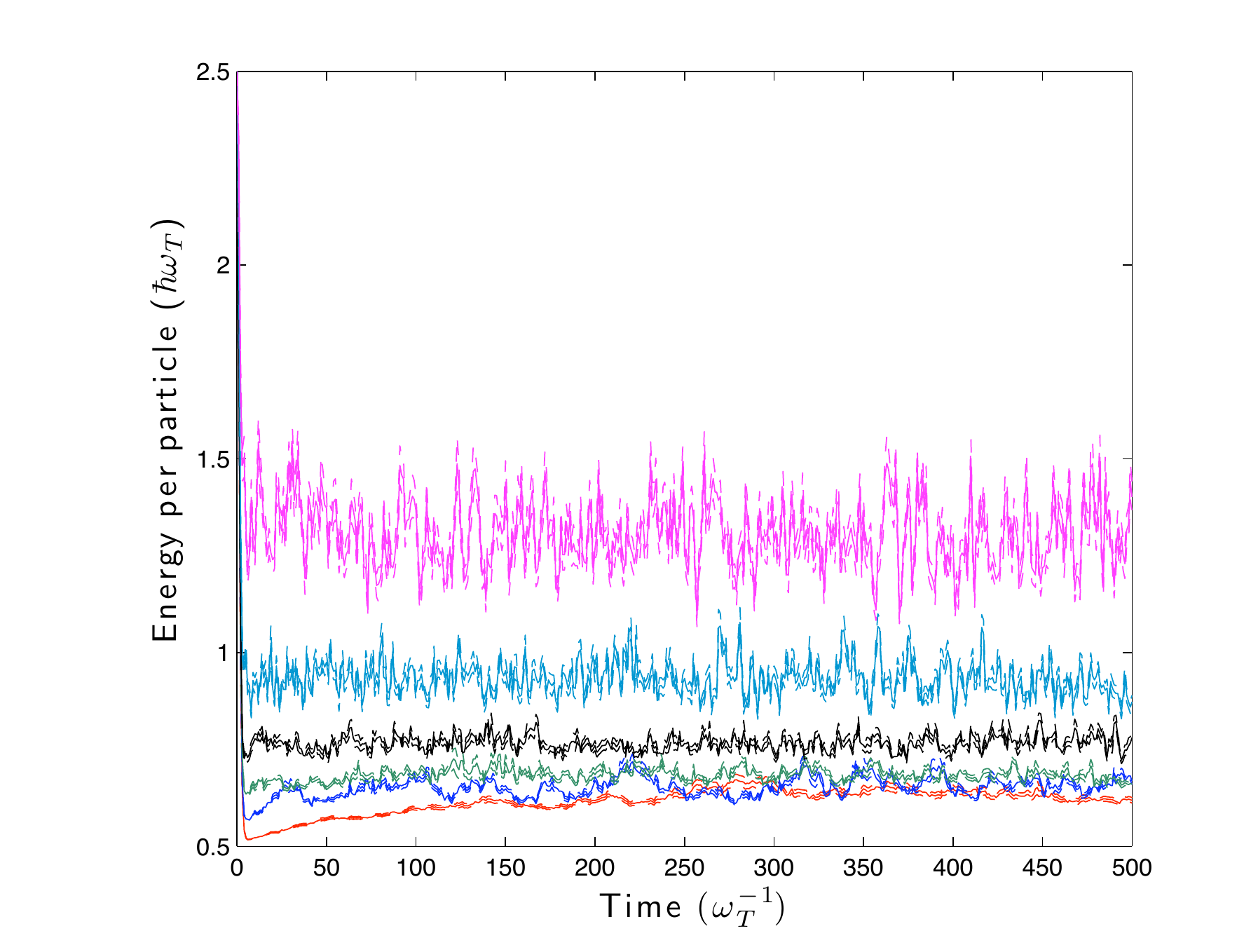}
\caption{\label{change_alpha} (Color online) Plots of the energy for simulations of 500 paths. Simulations were for a normalized Gaussian function centered at $x = 2.0$. Parameters chosen were $c_1 = 2.0$, $w = 3000.0$, and $\eta = 6.0$. The measurement strength (in order from bottom to top) is $\tilde\alpha = $ 2.0 (red) , 10.0 (blue), 20.0 (green), 40.0 (black), 80.0 (cobalt), 160.0 (magenta). Full lines represent the mean, while dotted lines indicate the standard error. Notice that those simulations with $\tilde\alpha \leq 20.0$ converge to the same average steady-state energy.}
\end{figure}
Ultimately, we are interested in how the average energy of the system $\sE\left[ \left< x^2 \right> + \left< p^2 \right> \right]/2$ varies over time; in particular the steady state value for the energy. In the following analysis we judge the effectiveness of the control based upon how close the system is cooled to the ground state energy $\hbar \omega_T/2$ and the time taken for the system to reach a steady-state. The energy was calculated by finding a numerical solution to the stochastic Schr\" odinger equation corresponding to eq.~(\ref{rho_c_single}). The numerical integration was performed by using the open source software package xpdeint, which is a new version of the xmds package \cite{xmds}.

Simulations revealed three important features of the system. The first relates to the measurement strength parameter $\tilde{\alpha}$. For a sufficiently large $\tilde\alpha$ the final average steady state energy increases with increasing measurement strength. This is because the measurement has a greater backaction on the atom for a larger measurement strength, which corresponds to an increased heating rate. Thus, a small $\tilde\alpha$ is required for a low steady-state energy. A caveat, however, to choosing a small $\tilde\alpha$ is that less information is obtained about the system per unit time. This translates to an increase in the time it takes for the energy to reach a steady-state. For optimal control, one would like to balance these considerations by choosing an $\tilde\alpha$ that cools close to the ground state energy on a timescale much smaller than that of the experiment. However, an additional constraint is that the cooling does not continue to get better as $\tilde \alpha$ decreases. There is a threshold value $\tilde \alpha_c$, where for any $\tilde\alpha \leq \tilde\alpha_c$ the final average energy is the same as that for $\tilde \alpha_c$. Indeed, the only effect of decreasing $\tilde \alpha$ lower than $\tilde \alpha_c$ is that it takes longer before the energy reaches a steady-state. A demonstration of this dependence on $\tilde \alpha$ is shown in Figure~\ref{change_alpha}. If no other constraints are taken into consideration, then this suggests that $\tilde\alpha_c$ is in fact the optimal value for $\tilde\alpha$.

\begin{figure}[t!]
	\centering
	\includegraphics[scale=0.55]{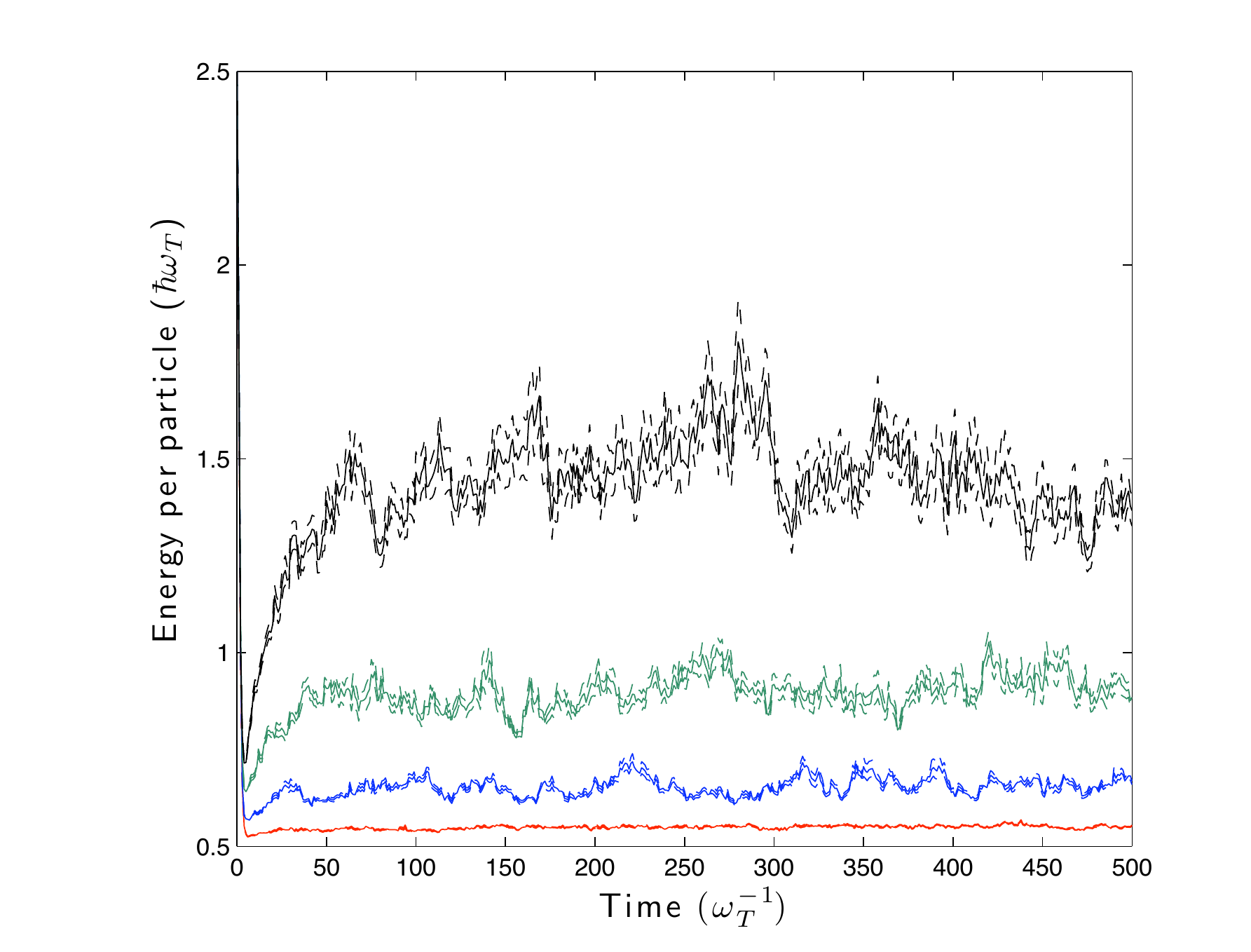}
\caption{\label{change_eta} (Color online) 500 path simulations of the energy for an initial normalized Gaussian state centered at $x = 2.0$, $w = 3000.0$, $\tilde\alpha = 10.0$, and $c_1 = 2.0$. The Lamb-Dicke parameter (in order from bottom to top) is $\eta = 4.0$ (red), 6.0 (blue), 8.0 (green), 10.0 (black). Full lines represent the mean, while dotted lines indicate the standard error. This plot shows that relatively small increases in $\eta$ result in large increases in the final average energy.
}
\end{figure}
The second point to consider is the dependence on the final steady state energy on the Lamb-Dicke parameter $\eta$. For small $\eta$, the energy imparted to the system during the scattering of the light is negligible. However, for large $\eta$ the centre of mass dynamics of the atom is greatly influenced. This introduces additional heating into the system. It was found that for larger values of $\eta$, the higher the final average energy (see Figure~\ref{change_eta}). Furthermore, while the final average energy scales roughly linearly with $\tilde\alpha$, it seems that it scales at a greater rate for $\eta$. Preliminary investigations seem to indicate that this scaling is quadratic. However, a more thorough study is required in order to draw a more accurate conclusion on this scaling. For typical trap frequencies ($\omega \sim 0.1$ Hz - 1 kHz), we can see that for a rubidium atom, $\eta$ will take values somewhere between 1 and 400. These results, coupled with the limited ability to reduce the height of the plateau by decreasing $\tilde\alpha$, indicate that this measurement and control scheme may only effectively cool a trapped atom for strong trapping potentials. 

\begin{figure}[t!]
\centering
\includegraphics[scale=0.55]{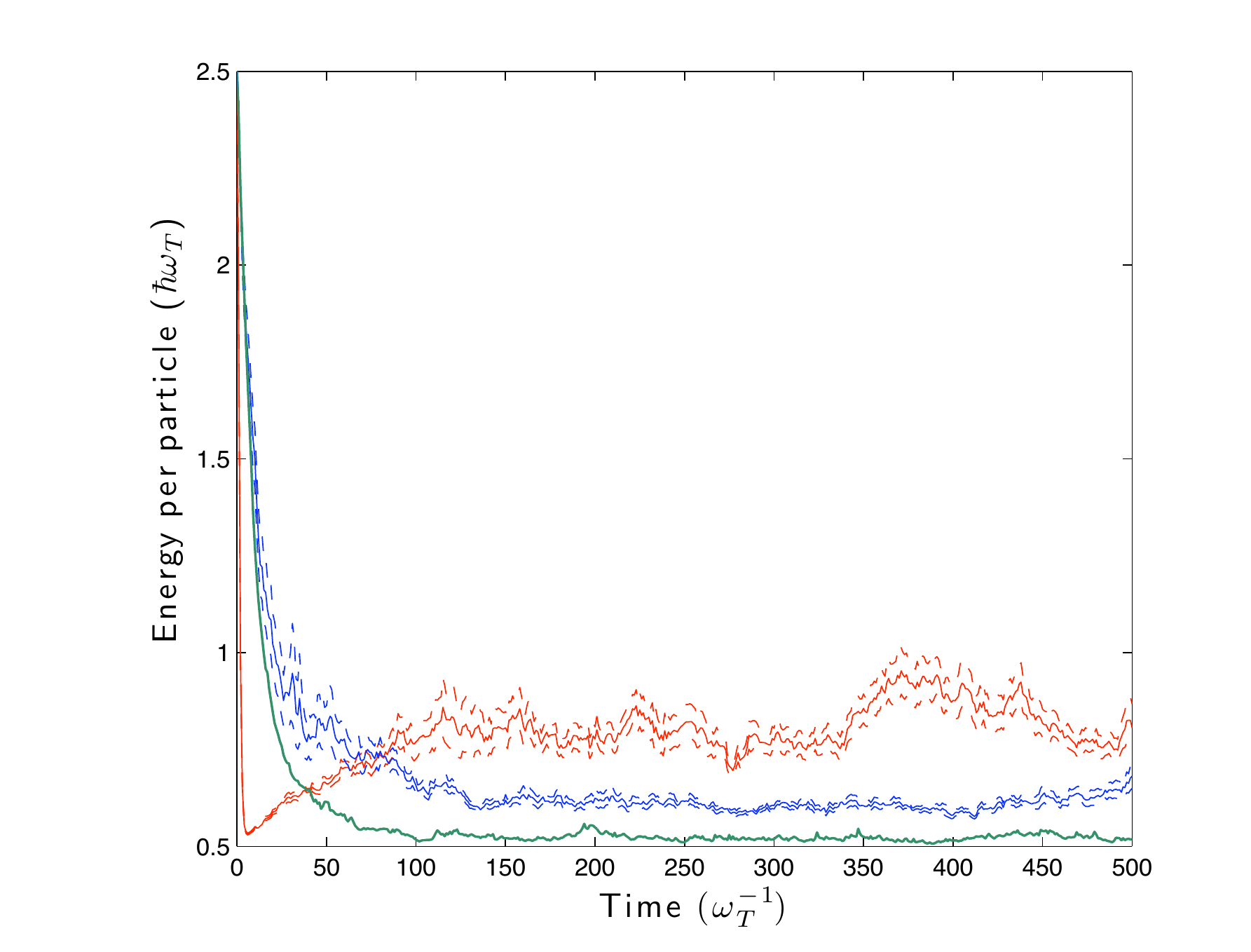}
\caption{\label{higher_controls} (Color online) 100 path simulations of the energy in the single particle approximation with feedback terms proportional to $x$ (red/highest), $x$ and $x^2$ (blue/middle), and $x, x^2, x^3$ (green/lowest). The parameters chosen were $w = 3000.0$, $\tilde\alpha = 2.0$, and $\eta = 8.0$. Full lines represent the mean, while dotted lines indicate the standard error. The initial states chosen were normalized Gaussian functions centered at $x = 2.0$. The constants of proportionality for each feedback term were set to 2.0.}
\end{figure} 
The third interesting feature involves a consideration of the effect of adding additional control terms of the form (\ref{control_H}) to the Hamiltonian. As demonstrated in Figure~\ref{higher_controls}, additional feedback terms increase the effectiveness of the control. In fact this is only true for larger values of $\eta$; if $\eta < 1$ then the higher order controls have little effect on the final average energy. This is precisely because for a large $\eta$ there is significant coupling between the different modes of the atom. This coupling is due to nonlinearities introduced by the trapped atom experiencing different electric fields at different positions in space. Hence energy can be removed from those higher order modes which are not directly affected by the feedback control. This is in contrast to a situation with small $\eta$, where higher order modes remain unaffected by the control of the lower order modes. Thus, despite the additional heating due to a large $\eta$, it may be possible to effectively cool the atom close to the ground state with the introduction of additional control terms.

\subsection{Mean-field approximation}
Although the single particle limit has its uses for qualitative understanding, multi-particle effects are important in many BEC phenomena. Indeed, the dynamics resulting from collisions between the atoms simply cannot be included in a single particle model. The next step towards modelling the above situation for a BEC with interactions is to consider the quantum filter (\ref{rho_c}) under the mean-field approximation
$$\phi(\xi) \approx \left< \hat{\psi}(\xi)\right>.$$
This yields the following Ito equation of motion for the mean-field wavefunction $\phi(\xi)$:
\begin{align}
	d\phi(\xi) 	&= \left(-iH_a(\xi) \phi(\xi) - \frac{\tilde{\alpha}}{2}\int d\kappa_x \, \tilde{\gamma}(\kappa_x)\right) \phi(\xi) d\tau \notag \\
			& \qquad + \tilde{\alpha}\phi(\xi)\int d\kappa_x \sqrt{\tilde{\gamma}(\kappa_x)}e^{-i \eta \kappa_x \xi}d\overline{W}^*(\kappa_x,\tau). \label{mean_field}
\end{align}

In contrast to the single atom limit, numerical simulations of eq.~(\ref{mean_field}) are not convergent. We conjecture that the difficulty stems from an inherent mismatch between the mean-field approximation and the type of measurement associated with phase-contrast imaging. This imaging gives information about the average density of the condensate, and is in effect a measurement of the number of atoms in the BEC. Thus over time, the continuous measurement projects the BEC state towards a number state. However, upon making the mean-field approximation it is assumed that the number variance is always that of a coherent state. This incongruity with the measurement may explain why simulations of the quantum filter under this semiclassical approximation give non-physical results.  

\section{Summary and conclusions}
This paper has investigated the prospect of generating a stable spatial mode for a trapped BEC by using feedback control to cool the condensate close to the ground state. In particular, the state of the BEC was estimated using a realistic measurement scheme based upon non-destructive imaging. We presented the full Hamiltonian for the system and bath, and derived a stochastic master equation for the conditional density matrix of the system. 

This SME was simulated in the single atom limit, and it was shown that a feedback scheme which adjusted the trap minimum would bring the atom to a steady-state energy. However, the precise value for this final energy and the time taken to bring the atom to that energy depends upon the measurement strength, Lamb-Dicke parameter (ratio of scattered photon's kinetic energy to atom's kinetic energy) and feedback strength. Simulations revealed that decreasing the measurement strength decreased the average steady state energy, but only up to a point. Below a certain critical value the steady state did not decrease; only the time taken to reach a steady state increased. A more problematic issue is the additional heating introduced into the system for values of the Lamb-Dicke parameter corresponding to some weakly trapped BEC experiments. However, it was demonstrated that much of this heating could be offset by the introduction of additional control terms proportional to higher powers of $x$. 

The derived SME was also simulated under the mean-field approximation, for a condensate with no interaction energy. These simulations did not converge.  We hypothesise that this lack of convergence is due the measurement projecting the state to an eigenstate that is no longer coherent, which is a dynamical effect at odds with the approximation itself.  We are currently investigating the behaviour of this filter under a number-conserving semiclassical approximation.  Ultimately, however, a definite answer to this question may only be obtained via a full field calculation. Such full field calculations may be possible using stochastic techniques that are also currently under investigation \cite{Hush_Carvalho_Hope_2009}.

\begin{acknowledgments}
The authors would like to acknowledge the help of Graham Dennis with the simulations.  This work is supported by the Australian Research Council Centre of Excellence for Quantum-Atom Optics and the National Computational Infrastructure National Facility.
Facility. 
\end{acknowledgments}

\appendix

\section{Derivation of Ito stochastic differential equation for unitary evolution \label{appendix_derivation}}
In this appendix we present a derivation of eq.~(\ref{dU}) from the total Hamiltonian (\ref{H_tot}).  The key physical approximation involved is that the optical field acts as a Markovian reservoir.  This requires a modal restriction of the atomic field, which is equivalent to allowing the possibility of a light pulse to travel away from the BEC and have no further interaction with it.  Although it is possible to make reservoir approximations for the light when a finite number of modes all interact with the same reservoir, we shall simplify our geometry by assuming that the magnetic potential is sufficiently tight in the $z$ direction such that the BEC is highly restricted in this dimension. This gives a `pancake'-shaped condensate, and restricts the BEC to the occupation of a single mode $z$ direction. That is
\begin{equation}
	\hat{\psi}_g(\bm{r}) \approx g(z)\hat{\psi}_g(\bm{x}), \quad \hat{\psi}_e(\bm{r}) \approx g(z) e^{ik_0 z}\hat{\psi}_e(\bm{x}),
\end{equation}
where $\bm{x} = (x,y)$. $\hat{\psi}_i(\bm{x})$ is interpreted as the two dimensional field operator for the $i$th state at position $\bm{x}$. The inclusion of the factor $e^{ik_0 z}$ represents the phase shift an atom receives upon being excited by a photon of momentum on the order of $k_0 = \omega_0/c$. Under this approximation
\begin{align}
	\hat{H}_I 	&= -\int dz \, |g(z)|^2 \int d^2\bm{x} \left[ \bm{d}_{ge} \cdot \hat{\bm{E}}(\bm{r},t)\right]  \notag \\
			& \quad \times \left(\hat{\psi}^\dag_e(\bm{x})\hat{\psi}_g(\bm{x})e^{-ik_0 z} + \text{h.c}\right). 
\end{align}
It is reasonable to assume that the density profile of the condensate in the $z$ direction is a normalised Gaussian of width $z_0$. Specifically
\begin{equation}
	|g(z)|^2 = \frac{1}{z_0\sqrt{\pi} }e^{-z^2/z_0^2}.
\end{equation}

Notice that before the dimensionality of the condensate was restricted there was one mode of light per mode of the condensate. After restriction, there are two infinite dimensions of radiation for every mode of BEC. This allows us to treat the $z$ component of the electromagnetic field as a reservoir. This is most simply done by first transforming the integral over $k_z$ into an integral over frequency $\omega$. The frequency is related to the wave number in the $z$ direction by the relationship
\begin{equation}
	\omega(k_z) = c\sqrt{|\bm{k}|^2 + k_z^2},
\end{equation}
for $\bm{k} = (k_x,k_y)$. This change of variables gives 
\begin{align}
	\hat{H}_I &= -i \int d^2\bm{k} \int_{c|\bm{k}|}^\infty d\omega \sqrt{\frac{\hbar \omega^2}{2(2\pi)^3 c^2 k_z \eo}} \left[ \bm{d}_{ge} \cdot \bm{\varepsilon}(\bm{k},\omega)\right] \notag \\
		& \times \left(\hat{L}^\dag(\bm{k})\hat{a}(\bm{k},\omega)G(k_z - k_0) + \hat{L}(\bm{k})\hat{a}(\bm{k},\omega)G(k_z + k_0) \right. \notag \\
		& \left. - \hat{L}^\dag(\bm{k})\hat{a}^\dag(\bm{k},\omega)G(k_z + k_0) - \hat{L}(\bm{k})\hat{a}^\dag(\bm{k},\omega)G(k_z - k_0) \right)
\end{align} 
where
\begin{align}
	\hat{L}(\bm{k}) &= \int d^2\bm{x} \, \hat{\psi}^\dag_g(\bm{x}) \hat{\psi}_e(\bm{x}) e^{-i\bm{k} \cdot \bm{x}} \label{L}
\end{align}
and
\begin{align}
	G(k_z) 	&= \frac{1}{\sqrt{2\pi}}\int dz |g(z)|^2 e^{-ik_0 z} e^{-ik_z z} \notag \\
			&= \exp\left[ -\frac{1}{4}z_0^2\left(k_z - k_0\right)^2\right]. 
\end{align}
The annihilation operators have been undergone the rescaling $ \hat{a}(\bm{p}) \to \sqrt{c^2 k_z/\omega} \, \hat{a}(\bm{k},\omega)$ to ensure that the commutation relation 
\begin{equation}
	[\hat{a}(\bm{k},\omega),\hat{a}^\dag(\bm{k}',\omega')] = \delta^2(\bm{k} - \bm{k'})\delta(\omega - \omega')
\end{equation}
is preserved. Note that even though $\pm k_z$ give the same $\omega$, the operators $\hat{a}(\bm{k},k_z)$ and $\hat{a}(\bm{k},-k_z)$ act on different Hilbert spaces. However, given most of the scattering will be in the positive $z$ direction, we have ignored the contribution due to $\hat{a}(\bm{k},-k_z)$. The Hamiltonian for the electromagnetic field becomes
\begin{equation}
	H_B = \int d^2 \bm{k} \int d\omega \, \hbar \omega \, \hat{a}^\dag(\bm{k},\omega)\hat{a}(\bm{k},\omega).
\end{equation}

We move into the interaction picture with the unitary transformation
\begin{equation}
	\hat{U}_{I}(t) = \exp\left(-\frac{i}{\hbar}\left(\hat{H}_B + \hbar\omega_L\int d^2\bm{x} \,\hat{\psi}_e^\dag(\bm{x})\hat{\psi}_e(\bm{x})\right)t\right), \label{U_I}
\end{equation}
where $\omega_L$ is the optical frequency of the laser. This transforms the operators as follows:
\begin{equation}
	\hat{a}_\pm(\bm{k},\omega) \to \hat{a}_\pm(\bm{k},\omega)e^{-i\omega t}, \quad \hat{L}(\bm{x}) \to \hat{L}(\bm{x})e^{-i\omega_L t}.
\end{equation}
Now the coupling of the atoms to the electromagnetic field will occur predominantly in a narrow frequency range $\omega_0 - \theta < \omega < \omega_0 + \theta$ for some $\theta \ll \omega_0$. The rotating wave approximation can thus be used to neglect those terms which are rotating quickly. Hence 
\begin{align}
	\hat{H}_I &= -i \int_\Omega d^2\bm{k} \int_{\omega_0 - \theta}^{\omega_0 + \theta} d\omega \, \kappa(\bm{k},\omega)\left(\hat{L}^\dag(\bm{x})\hat{a}(\bm{k},\omega)e^{- i(\omega - \omega_L)t} \right. \notag \\
		& \qquad \left.  - \hat{L}(\bm{x})\hat{a}^\dag(\bm{k},\omega)e^{ i(\omega - \omega_L)t} \right).
\end{align} 
where we have defined the strength of coupling between the system and photon bath as
\begin{equation}
	\kappa(\bm{k},\omega) = \sqrt{\frac{\omega^2d_{ge}^2}{2(2\pi)^3 c^2 k_z \hbar \eo}} \left[ \hat{\bm{d}}_{ge} \cdot \bm{\varepsilon}(\bm{k},\omega)\right] G(k_z - k_0).
\end{equation}
The domain of integration over $\bm{k}$ has been restricted to $\Omega = \{ \bm{k} \colon c|\bm{k}| < \omega_0\}$ to ensure that $\omega$ is never complex. It is assumed that the coupling strength is roughly constant in frequency space around the resonant frequency. That is, $\kappa(\bm{k},\omega) \approx \kappa(\bm{k},\omega_0)$. Furthermore, let
\begin{equation}
	\hat{a}^{(\theta)}(\bm{k},t) = \frac{1}{\sqrt{2\pi}}\int_{\omega_0 - \theta}^{\omega_0 + \theta} d\omega \, \hat{a}(\bm{k},\omega) e^{-i(\omega-\omega_L)t}.
\end{equation}
Then
\begin{equation}
	\hat{H}_I = i\hbar \int_\Omega d^2\bm{k} \, \sqrt{2\pi}\kappa(\bm{k},\omega_0)\left(\hat{L}(\bm{k})\hat{a}^{(\theta)\dag}(\bm{k},t) - h.c.\right).
\end{equation}

So far no direct assumptions have been made about the nature of the electromagnetic field coupled to the system. This field can be well approximated as a classical light field with quantum vacuum fluctuations. Make the replacement ${\hat{a}^{(\theta)}(\bm{k},t) \to \hat{a}^{(\theta)}(\bm{k},t) + f(\bm{k},t)}$, and approximate the entire bath as a vacuum state. The interaction Hamiltonian becomes
\begin{align}
	\hat{H}_I &= i\hbar \int_\Omega d^2\bm{k} \, \sqrt{2\pi}\kappa(\bm{k},\omega_0) \left\{\left(\hat{L}(\bm{k})\hat{a}^{(\theta)\dag}(\bm{k},t) -h.c.\right)\right. \notag \\ 
		& \qquad \left.+ \left(\hat{L}(\bm{k})f^*(\bm{k},t) - \hat{L}^\dag(\bm{k})f(\bm{k},t)\right)\right\}.  \label{H_I_approx}
\end{align}

In the interaction picture, the evolution of the unitary operator $\hat{U}_I$ is
\begin{align}
	\frac{d}{dt}\hat{U}_I(t) &= -\frac{i}{\hbar}\left( \hat{H}_I + \int d^2\bm{x} \,\hat{\psi}_g^\dag(\bm{x})H_a(\bm{x})\hat{\psi}_g(\bm{x}) \right. \notag \\
						& \, \left. + \int d^2\bm{x} \,\hat{\psi}_e^\dag(\bm{x})\left(H_a(\bm{x}) + \hbar \Delta \right)\hat{\psi}_e(\bm{x}) \right) \hat{U}_I(t),
\end{align}
where $\Delta = \omega_0 - \omega_L$. Substituting (\ref{H_I_approx}) into this expression gives
\begin{align}
	\frac{d}{dt}\hat{U}^{(\theta)}(t) &= \left\{-\frac{i}{\hbar}\hat{H} +  \int_\Omega d^2\bm{k} \, \sqrt{2\pi}\kappa(\bm{k},\omega_0) \right. \notag \\
		& \, \times \left. \left(\hat{L}(\bm{k})\hat{a}^{(\theta)\dag}(\bm{k},t)  - h.c. \right)\right\}\hat{U}^{(\theta)}(t),  \label{dU_theta} 
\end{align}
where
\begin{align}
	\hat{H} 	&= \int d^2\bm{x}\, \hat{\psi}_g^\dag(\bm{x})H_a(\bm{x})\hat{\psi}_g(\bm{x}) \notag \\
			& \quad + \int d^2\bm{x} \, \hat{\psi}_e^\dag(\bm{x})\left(H_a(\bm{x}) + \hbar \Delta \right)\hat{\psi}_e(\bm{x}) \notag \\
			& \quad + i\hbar\int_\Omega d^2 \bm{k} \, \sqrt{2\pi}\kappa(\bm{k},\omega) \left( \hat{L}(\bm{k})f^*(\bm{k},t) - h.c. \right).
\end{align}
The superscript $(\theta)$ has been used simply to highlight the dependence of the unitary on $\theta$. It is reasonable to assume that the BEC is much smaller in the $x$ and $y$ directions than the spatial size of the coherent beam. This allows the laser to be adequately approximated as a plane wave propagating in the $z$ direction with frequency $\omega_L$. Thus $f(\bm{k},t) \approx \sqrt{F_0} \delta^2(\bm{k})$, where $F_0$ is the photon number flux of the laser. Under this approximation the Hamiltonian $\hat{H}$ reduces to eq.~(\ref{H_f}).

Now, note that as $\theta \to \infty$,
$$\hat{a}^{(\theta)}(\bm{k},t) \to \hat{a}(\bm{k},t) = \frac{1}{\sqrt{2\pi}}\int_{-\infty}^\infty \hat{a}(\bm{k},\omega)e^{-i(\omega-\omega_0)t}.$$
In vacuum $\left< \hat{a}(\bm{k},t) \right> = 0$ and $\left< \hat{a}^\dag(\bm{k},t)\hat{a}(\bm{k},s)\right> = \delta(t - s)$, and can therefore be identified as quantum white noise. Given that the coupling is weak, in the sense that $\kappa(\bm{k},\omega_0) \ll 1$, eq.~(\ref{dU_theta}) indicates that the timescale on which the system evolves will be slow. Since the frequency range of coupling is narrow, it can be assumed that $\theta \gg [\kappa(\bm{k},\omega_0)]^2$. Hence the system is well described by taking $\theta$ as practically infinite \cite{Gardiner_Zoller_2004}. The importance of this observation is that the quantum analogue of the Wong-Sakai theorem states that in the limit of $\hat{a}_\pm^{(\theta)}(\bm{k},t)$ approaching quantum noise (i.e. $\theta \to \infty$) the solution to eq.~(\ref{dU_theta}) approaches the solution to the Ito quantum stochastic differential equation (QSDE) 
 \begin{align}
	&d\hat{U}(t) = \left\{\left(-\frac{i}{\hbar}\hat{H} - \half \int_\Omega d^2 \bm{k} \, \gamma(\bm{k}) \hat{L}^\dag(\bm{k})\hat{L}(\bm{k})\right) dt  \right.  \notag \\
			& \left.+ \int_\Omega d^2\bm{k} \, \sqrt{\gamma(\bm{k})} \left(\hat{L}(\bm{k}) d\hat{A}^\dag(\bm{k},t)- \hat{L}^\dag(\bm{k}) d\hat{A}(\bm{k},t)\right)\right\}\hat{U}(t). \label{dU_k}
\end{align}
where we have defined
\begin{equation}
	\gamma(\bm{k}) = 2\pi \int_\Omega d^2\bm{k} \, [ \kappa(\bm{k},\omega_0)]^2,
\end{equation}
and $d\hat{A}(\bm{k},t)$ is the quantum Wiener increment. It satisfies the property $d\hat{A}(\bm{k},t)d\hat{A}^\dag(\bm{k}',t) = \delta(\bm{k} - \bm{k}') dt$. For an heuristic development of this theorem, see \cite{van_Handel_2005}. A more rigorous treatment can be found in \cite{Accardi_1990, Gough_2005}. Thus, to a good approximation, the unitary for the total system is given by the QSDE (\ref{dU_k}). Using the definition of $\hat{L}$ given in eq.~(\ref{L}) and 
\begin{equation}
	d\hat{A}(\bm{x},t) = \frac{1}{\sqrt{2\pi}}\int d^2 \bm{k} \hat{A}(\bm{k},t) e^{-i\bm{k}\cdot \bm{x}}
\end{equation}

QSDE (\ref{dU_k}) can be written in the form
\begin{align}
	& d\hat{U}(t) = \left\{\left(-\frac{i}{\hbar}\hat{H} - \half\int d^2\bm{x} \, \hat{M}^\dag(\bm{x})\hat{M}(\bm{x}) \right) dt  \right.  \notag \\
			& \quad \left.+ \int d^2\bm{x} \, \left(\hat{M}(\bm{x}) d\hat{A}^\dag(\bm{x},t) - \hat{M}^\dag(\bm{x}) d\hat{A}(\bm{x},t)\right)\right\}\hat{U}(t), \label{dU_2}
\end{align}
where
\begin{align}
&\hat{H} 	= \int d^2\bm{x} \, \hat{\psi}_g^\dag(\bm{x})H_a(\bm{x})\hat{\psi}_g(\bm{x}) \notag \\
		& \, + \int d^2 \bm{x} \, \hat{\psi}_e^\dag(\bm{x})\left(H_a(\bm{x}) + \hbar \Delta \right)\hat{\psi}_e(\bm{x}) \notag\\
			& \, + i \sqrt{\frac{F_0 k_0^2 \hbar}{2(2\pi)^3 \eo}}  [ \bm{d}_{ge} \cdot \bm{\epsilon}(\bm{0})]\int d^2\bm{x} \, \left( \hat{\psi}^\dag_g(\bm{x})\hat{\psi}_e(\bm{x}) - h.c. \right), \label{H_f_2}
\end{align}
and
\begin{align}
	\hat{M}(\bm{x})		&= \int d^2\bm{x}' \hat{\psi}_g^\dag(\bm{x}') \hat{\psi}_e(\bm{x}')\Gamma(\bm{x} - \bm{x}'), \notag \\
	\Gamma(\bm{x})	&= \int_\Omega d^2 \bm{k} \, \sqrt{\frac{\gamma(\bm{k})}{2\pi}}e^{i\bm{k} \cdot \bm{x}}, \notag \\
	\gamma(\bm{k})	&= \frac{k_0^2}{8\hbar \eo} [ \bm{d}_{ge} \cdot \bm{\epsilon}]^2 \frac{e^{ -z_0^2 (k_0 - k_z)^2/2}}{\sqrt{k_0 ^2 - k_x^2 - k_y^2}}, \notag
\end{align}
for $d\hat{A}(\bm{x},t)d\hat{A}^\dag(\bm{x}.,t) = \delta^2(\bm{x} - \bm{x}')dt$.

As a final note, a `cigar'-shaped condensate can be considered by assuming the BEC is tightly confined in both the $z$ and $y$ directions. In this case
\begin{equation}
	\hat{\psi}_g(\bm{r}) \approx g(z)h(y) \hat{\psi}_g(x), \quad \hat{\psi}_e(\bm{r}) \approx g(z)h(y) e^{ik_0 z}\hat{\psi}_e(x),
\end{equation}
where $\hat{\psi}_i(x)$ is the one dimensional field operator for the $i$th state at position $x$ and 
\begin{equation}
	 |h(y)|^2 = \frac{1}{z_0\sqrt{\pi}}e^{-y^2/z_0^2}.
\end{equation}
For simplicity we have assumed that the density profiles in the $y$ and $z$ directions are both normalised Gaussians of identical width $z_0$. A similar argument to that given above leads to the evolution of the one dimensional unitary
 \begin{align}
	d\hat{U}(t) 	&= \left\{\left(-\frac{i}{\hbar}\hat{H} - \half \int_\Omega d^2 \bm{k} \, 2\pi [\kappa_1(\bm{k},\omega_0)]^2 \hat{L}^\dag(k_x)\hat{L}(k_x)\right) dt  \right.  \notag \\
			& \qquad \left.+ \int_\Omega d^2\bm{k} \, \sqrt{2\pi}\kappa_1(\bm{k},\omega_0) \left(\hat{L}(k_x) d\hat{A}^\dag(\bm{k},t) \right. \right. \notag \\
			&  \left. \left. \qquad \qquad  - \hat{L}^\dag(k_x) d\hat{A}(\bm{k},t)\right)\right\}\hat{U}(t). \label{dU_kone}
\end{align}
where the coupling constant is now given as
\begin{equation}
	\kappa_1(\bm{k},\omega) = \sqrt{\frac{\omega^2d_{ge}^2}{2(2\pi)^3 c^2 k_z \hbar \eo}} \left[ \hat{\bm{d}}_{ge} \cdot \bm{\varepsilon}(\bm{k},\omega)\right] G(k_z - k_0)H(k_y),
\end{equation}
for 
\begin{align}
	H(k_y) 	&= \frac{1}{\sqrt{2\pi}}\int dy |g(y)|^2 e^{-ik_y y},
\end{align}
and
\begin{align}
	\hat{L}(k_x) &= \int dx \hat{\psi}^\dag_g(x) \hat{\psi}_e(x) e^{-ik_x x}.
\end{align}
In this form, it is possible to integrate out the explicit dependence on $k_y$. Define 
\begin{equation}
	\gamma(k_x) = 2\pi \int_{-\sqrt{k_0^2 - k_x^2}}^{\sqrt{k_0^2 - k_x^2}} dk_y\, [ \kappa_1(\bm{k},\omega_0)]^2,
\end{equation}
Also consider the one dimensional quantum Wiener increment $\hat{A}(k_x,t)$, which satisfies the relation $dA(k_x,t) dA^\dag(k_x',t) = \delta(k_x - k_x') dt$. So
\begin{align*}
	&\left(\int d^2 \bm{k} \sqrt{2\pi} \kappa_1(\bm{k},\omega_0) d\hat{A}(\bm{k},t)\right)\left(\int d^2 \bm{k'} \sqrt{2\pi} \kappa_1(\bm{k'},\omega_0) d\hat{A}(\bm{k'},t)\right)^\dag 	\\
		&=  \int d^2 \bm{k}\, 2\pi [\kappa_1(\bm{k},\omega_0)]^2 dt \\
		&= \int_{-k_0}^{k_0} dk_x \gamma(k_x)dt \\
		&= \left(\int_{-k_0}^{k_0} dk_x \sqrt{\gamma(k_x)}d\hat{A}(k_x,t)\right)\left(\int_{-k_0}^{k_0} dk_x' \sqrt{\gamma(k_x')}d\hat{A}(k_x',t)\right)^\dag.
\end{align*}
This shows that eq.~(\ref{dU_k}) obeys the same statistics as, and is therefore equivalent to, 
 \begin{align}
	&d\hat{U}(t) 	= \left\{\left(-\frac{i}{\hbar}\hat{H} - \half\int_{-k_0}^{k_0} dk_x \gamma(k_x) \hat{L}^\dag(k_x)\hat{L}(k_x)\right) dt  \right.  \notag \\
			& \left.  + \int_{-k_0}^{k_0} dk_x\, \sqrt{\gamma(k_x)} \left(\hat{L}(k_x) d\hat{A}^\dag(k_x,t) - \hat{L}^\dag(k_x) d\hat{A}(k_x,t)\right)\right\}\hat{U}(t). \label{dU_k_1}
\end{align}
As before, this can be rewritten to give eq.~(\ref{dU}).
\newpage 
\bibliography{single_atom_paper_final}

\end{document}